\documentclass[]{aastex}
\usepackage{psfig}

\newcommand\pp     {$\pm$}
\newcommand\pers     {s$^{-1}$}
\newcommand\micros  {$\mu$s}

\def\degr{\hbox{$^\circ$}}

\righthead{Burst oscillations in X 1658--298}
\slugcomment{Accepted for publication in ApJ Letters: 21 December, 2000}

\begin{document}

\title{Discovery of nearly coherent oscillations with a frequency of
$\sim$567 Hz during type-I X-ray bursts of the X-ray transient and
eclipsing binary X 1658--298}

\author{Rudy Wijnands\altaffilmark{1,4}, Tod Strohmayer\altaffilmark{2},
\& Luc\'\i a M. Franco\altaffilmark{3}}

\altaffiltext{1}{Center for Space Research, Massachusetts Institute of
Technology, 77 Massachusetts Avenue, Cambridge, MA 02139-4307, USA;
rudy@space.mit.edu; Chandra Fellow}

\altaffiltext{2}{Laboratory for High Energy Astrophysics, Goddard Space Flight
Center, Greenbelt, MD 20771}

\altaffiltext{3}{University of Chicago, 5640 S. Ellis Ave., Chicago IL 60637}

\altaffiltext{4}{Chandra Fellow}

\begin{abstract}
We report the discovery of nearly coherent oscillations with a
frequency of $\sim 567$ Hz during type-I X-ray bursts from the X-ray
transient and eclipsing binary X 1658--298. If these oscillations are
directly related to the neutron star rotation then the spin period of
the neutron star in X 1658--298 is $\sim$1.8 ms.  The oscillations can
be present during the rise or decay phase of the bursts. Oscillations
during the decay phase of the bursts show an increase in the frequency
of $\sim$0.5 to 1 Hz. However, in one particular burst the
oscillations reappear at the end of the decay phase at about 571.5
Hz. This represents an increase in oscillation frequency of about 5 Hz
which is the largest frequency change seen so far in a burst
oscillation. It is unclear if such a large change can be accommodated
by present models used to explain the frequency evolution of the
oscillations.  The oscillations at 571.5 Hz are unusually soft
compared to the oscillations found at 567 Hz.  We also observed
several bursts during which the oscillations are detected at much
lower significance or not at all.  Most of these bursts happen during
periods of X-ray dipping behavior, suggesting that the X-ray dipping
might decrease the amplitude of the oscillations (although several
complications exist with this simple picture).  We discuss our
discovery in the framework of the neutron star spin interpretation.

\end{abstract}


\section{Introduction \label{intro}}
 
Of the more than fifty low-mass X-ray binaries (LMXBs) that exhibit
type-I X-ray bursts only in six\footnote{After submission of our paper
NCOs were also reported in 4U 1916--053 (Galloway et al. 2000), 4U
1608--52 (Chakrabarty et al. 2000), and possibly SAX J1808.4--3658 (in
't Zand et al. 2000) bringing to ten the number of systems now known
to exhibit NCOs.}  systems have nearly coherent oscillations (NCOs)
during X-ray bursts been detected (Strohmayer et al. 1996, 1997b,
1998a; Smith, Morgan, \& Bradt 1997; Zhang et al. 1998; Markwardt,
Strohmayer, \& Swank 1999; these oscillations are also called 'burst
oscillations'). Their high coherence (e.g., Strohmayer \& Markwardt
1999; Muno et al. 2000), their strength (Strohmayer, Zhang, \& Swank
1997a; Strohmayer et al. 1998a), and their frequency stability over
many years (Strohmayer et al. 1998b) suggest that they are due to the
spin of the neutron star. Their observed frequencies (300--600 Hz) are
also consistent with the spin periods of other known rapidly rotating
neutron stars, for example, the millisecond radio pulsars, and the 401
Hz accreting X-ray pulsar SAX J1808.4--3658 (Wijnands \& van der Klis
1998).  Four of the burst oscillation systems are persistent sources
and can be studied extensively so that the properties of their bursts
(i.e., the NCOs) can be correlated with the overall behavior of the
source (i.e., the variations in the mass accretion rate), although
this has so far only been done for two systems (Muno et al. 2000;
Franco 2000; van Straaten et al. 2000). The other two sources are
transients and the NCOs can only be studied when these sources are in
outburst. Such outbursts occur infrequently and only a limited number
of X-ray bursts from them have been observed. Because of the limited
amount of published results, the phenomenology of how the properties
of NCOs relate to source state and mass accretion rate are not yet
well understood.

The LMXB X 1658--298 is an X-ray transient and was discovered by
Lewin, Hoffman, \& Doty (1976). They reported type-I X-ray bursts from
the system, and thus demonstrated that the compact object is a neutron
star. Cominsky \& Wood (1984, 1989) found that the source displays
deep X-ray dips and eclipses every $\sim$7.1 hours, which most likely
is the orbital period.  After an outburst in 1978 the source remained
in quiescence until April 1999, when it was again found to be active
(in 't Zand et al. 1999).  The {\it RXTE}/ASM light curve\footnote{The
ASM light curve of X 1658--298 can be found at
http://xte.mit.edu/ASM\_lc.html} shows that as of the time of
resubmission of this Letter (15 December 2000) the source is still
active.  An updated ephemeris for the orbital period was obtained by
Wachter, Smale, \& Bailyn (2000), using public TOO observations of X
1658--298 performed with the proportional counter array (PCA) on the
{\it Rossi X-ray Timing Explorer} ({\it RXTE}). We searched the
archival {\it RXTE} data (public TOO observations and proprietary data
which have become publicly available) of X 1658--298 for type-I X-ray
bursts and we then searched the bursts we found for the presence of
NCOs. Here, we report the discovery of NCOs at a frequency of
$\sim$567 Hz during X-ray bursts from X 1658--298. A preliminary
announcement of this discovery was already made by Wijnands,
Strohmayer, \& Franco (2000a).

\section{Observations, analysis, and results}

Since its reappearance, X 1658--298 has been observed on many
occasions with the {\it RXTE}/PCA, both as a result of proprietary
Cycle 4 observations and also via public TOO observations (Cycle 4 and
5). We searched all the currently public archival data for X-ray
bursts using the Standard 1 data mode. Besides this mode and the
Standard 2 mode, data were collected using an event mode with a time
resolution of 122 \micros~in 64 energy channels (E\_125us\_64M\_0\_1s;
covering the full {\it RXTE}/PCA energy range of 2--60 keV). These
event mode data were used to search for and study oscillations during
the type-I bursts.

We found 14 bursts (Tab.~\ref{tab:oscillations}), five of which
occurred during intervals of dipping activity (bursts 1, 5, 9, 10, and
12) and one occurred during an eclipse (burst 14) and only the end of
its decay phase could be studied. Of the bursts which occurred outside
the dips, one burst was only about half as bright as the others (burst
6).  Only during five bursts (bursts 1, 2, 3, 4, and 7) were all five
PCA detectors on and for the others only four (eight bursts; bursts 5,
6, 8, 9, 11, 12, 13, and 14) or three (one burst; burst 10) detectors
were active.

We searched the bursts for NCOs by using the $Z^{2}_{n}$ statistic
(i.e., the $Z^{2}_{1}$ statistic because NCOs are highly sinusoidal)
which was first used for analyzing the timing properties of X-ray
bursts by Strohmayer \& Markwardt (1999).  We computed $Z_1^2$
throughout the bursts using overlapping 2 s intervals with a new
interval defined every 1/4 s.  The $Z_1^2$ statistic has the same
statistical properties as a Leahy normalized power spectrum. Thus, for
a purely random Poisson process it is distributed as the $\chi^2$
distribution with 2 degrees of freedom.  We searched the frequency
range from 100--1200 Hz (for the range 2--22 keV) and discovered that
highly significant NCOs were present at a frequency of $\sim$567 Hz in
five bursts (Fig.~\ref{fig:oscillations_1}; bursts 2, 3, 4, 8, and
11).  In burst 9 the NCOs were also conclusively detected but at a
much lower significance level.  For a few of the remaining bursts we
have indications that the NCOs might also have been present, however,
these detections are not significant.  We quote 90\% confidence upper
limits on the strength of possible NCOs in these bursts
(Tab.~\ref{tab:oscillations}).

\begin{figure}[]
\begin{center}
\begin{tabular}{c}
\psfig{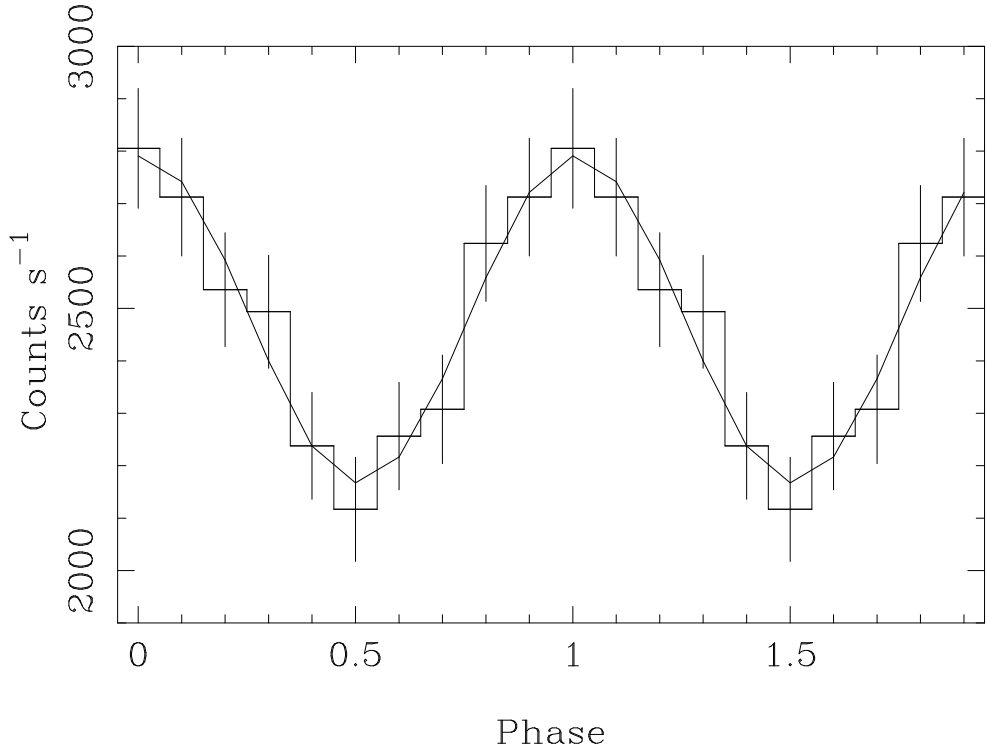}
\end{tabular}
\figcaption{Folded light curve in the energy range 2--22 keV
(including the persistent source count rate; for 5 PCUs) of the
oscillations observed during the tail of burst which started on 1999
April 10, 9:48 UTC (burst 3) at a frequency of $\sim$567 Hz. Two
periods are shown for clarity.
\label{fig:folded} }
\end{center}
\end{figure}

The strength of the NCOs is typically between 8 \% and 20 \% rms
amplitude in the energy range 2--60 keV (Tab.~\ref{tab:oscillations};
including the persistent source count rate which varied typically
100--150 counts \pers~PCU$^{-1}$) and the folded light curves of the
oscillations are very sinusoidal (e.g., Fig.~\ref{fig:folded}).  For
two of these bursts the NCOs were only present during the rising phase
of the bursts (e.g., Fig.~\ref{fig:oscillations_1}{\it a}) and for the
other four bursts (including burst 9) in the tail of the bursts (e.g.,
Fig.~\ref{fig:oscillations_1}{\it b}).  The properties of the method
used for the analysis of the NCOs and their transient nature causes
the broad distribution in frequency of the contours near the burst
rise in Figure~\ref{fig:oscillations_1}{\it a}.  It is still not clear
what determines the time or times during the bursts when oscillations
are present. We are in the process of analyzing the bursts behavior
and the source behavior in detail in order to determine if the
properties of the NCOs are correlated with any of the other burst
properties (e.g., the burst profile, episodes of photo-spheric radius
expansion) or with the properties of the source (e.g., mass accretion
rate). In a forthcoming paper we will discuss our results (Wijnands et
al. 2000b).

When the NCOs are found in the tail of the bursts, their frequency
increases slightly by 0.5--1 Hz (Fig.~\ref{fig:oscillations_1}{\it
b}). However, in one burst we found a very interesting evolution of
the oscillation frequency. The frequency during the burst decay phase
increased slightly by $\sim$1 Hz after which the oscillation died
away. However about 4 s later the oscillation reappeared but at a
frequency of 571.5 Hz (12.3\%\pp1.5\% rms amplitude; 2--60 keV;
Fig.~\ref{fig:oscillations_2} {\it top panel}). These NCOs near 571.5
Hz have similar peak power ($Z^{2}_{1} \sim 32$) as the ones earlier
on and the single trail probability of detecting such a power by
chance is 1.1$\times10^{-7}$. When taking into account a conservative
estimate of the number of trials involved (the frequency range
searched was 564--574 Hz for a 20 s interval after the burst peak),
this probability increases to 2.2$\times10^{-5}$. Including the number
of bursts (14) in our sample we find a probability of
3.1$\times10^{-4}$ that the signal is due to chance. However, when
taking into account only the six bursts which do show NCOs the
probability decreases to 1.3$\times10^{-4}$. We conclude from these
numbers that the signal we see at 571.5 Hz is real. Similar
oscillations were not present in the other bursts with typical
amplitude upper limits of 5\%--11\% rms (2--60 keV;
Tab.~\ref{tab:oscillations}).

The 5 Hz change in frequency represents the largest increase in any
burst oscillation reported to date. To study this behavior further we
decided to examine the energy dependence of the oscillations. We found
that the oscillations at 571.5 Hz are most clearly visible below 6 keV
(Fig.~\ref{fig:oscillations_2} {\it middle panel}), whereas when the
frequency is close to 567 Hz the oscillations are stronger above 6 keV
(Fig.~\ref{fig:oscillations_2} {\it bottom panel}). This demonstrates
that the oscillation had a different energy dependence at different
times during the burst.

\section{Discussion}

We have discovered highly significant nearly coherent oscillations at
a frequency of $\sim$567 Hz during five type I X-ray bursts observed
from the X-ray transient and eclipsing binary X 1658--298. This makes
this source the seventh source for which highly significant NCOs have
been found. The properties of these oscillations (their coherence,
their strengths, their long-term stability) make it likely that they
are directly related to the spin frequency of the neutron star (or its
first overtone as in 4U 1636--53; Miller 1999). In this
interpretation, the spin period of the neutron star in X 1658--298
would be $\sim$1.8 ms. This is well within the range of what has been
observed for the other sources which exhibit NCOs.

\begin{figure}[]
\begin{center}
\begin{tabular}{c}
\psfig{figure=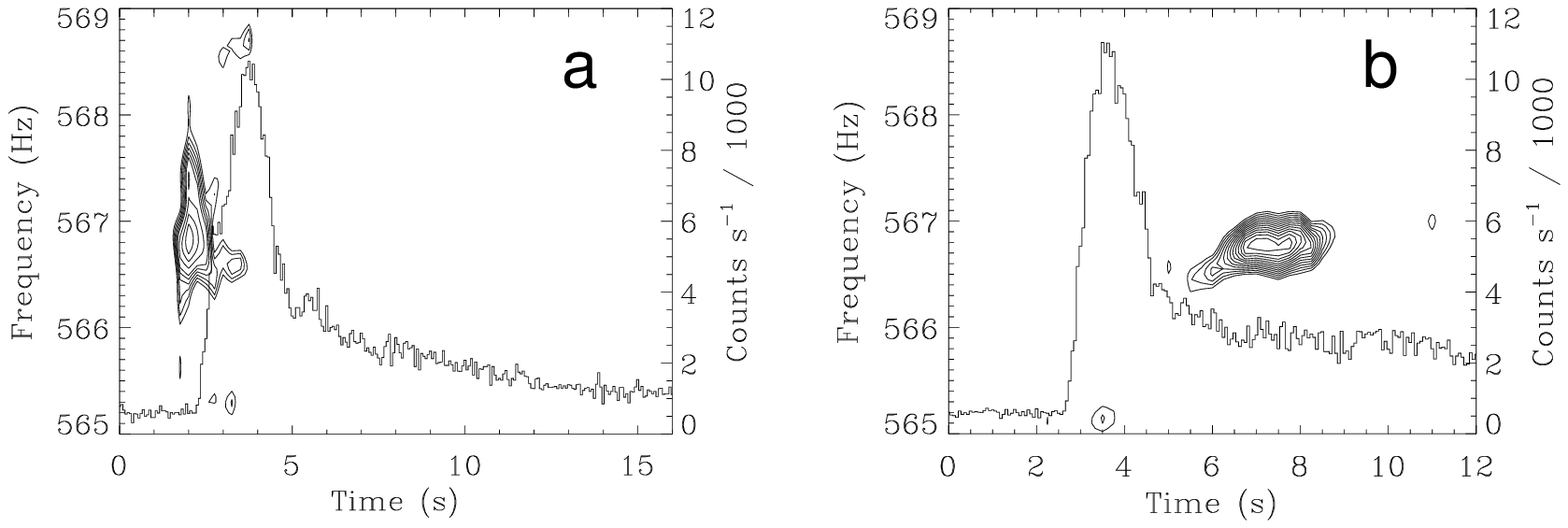,width=15cm}
\end{tabular}
\figcaption{The dynamical power spectra in the energy range 2--22 keV
of the bursts on 1999 April 9, 14:46 UTC (burst 2; {\it a}) and on
1999 April 10, 9:48 UTC (burst 3; {\it b}) which clearly show the
oscillations near 567 Hz.  The lowest contour plotted is $Z^{2}_{1}$ =
14, and the contours go up in steps of 2. The burst count rate
profiles (for 5 PCUs) are over-plotted. The times given correspond to
the start of the plots.
\label{fig:oscillations_1} }
\end{center}
\end{figure}

\begin{figure}[]
\begin{center}
\begin{tabular}{c}
\psfig{figure=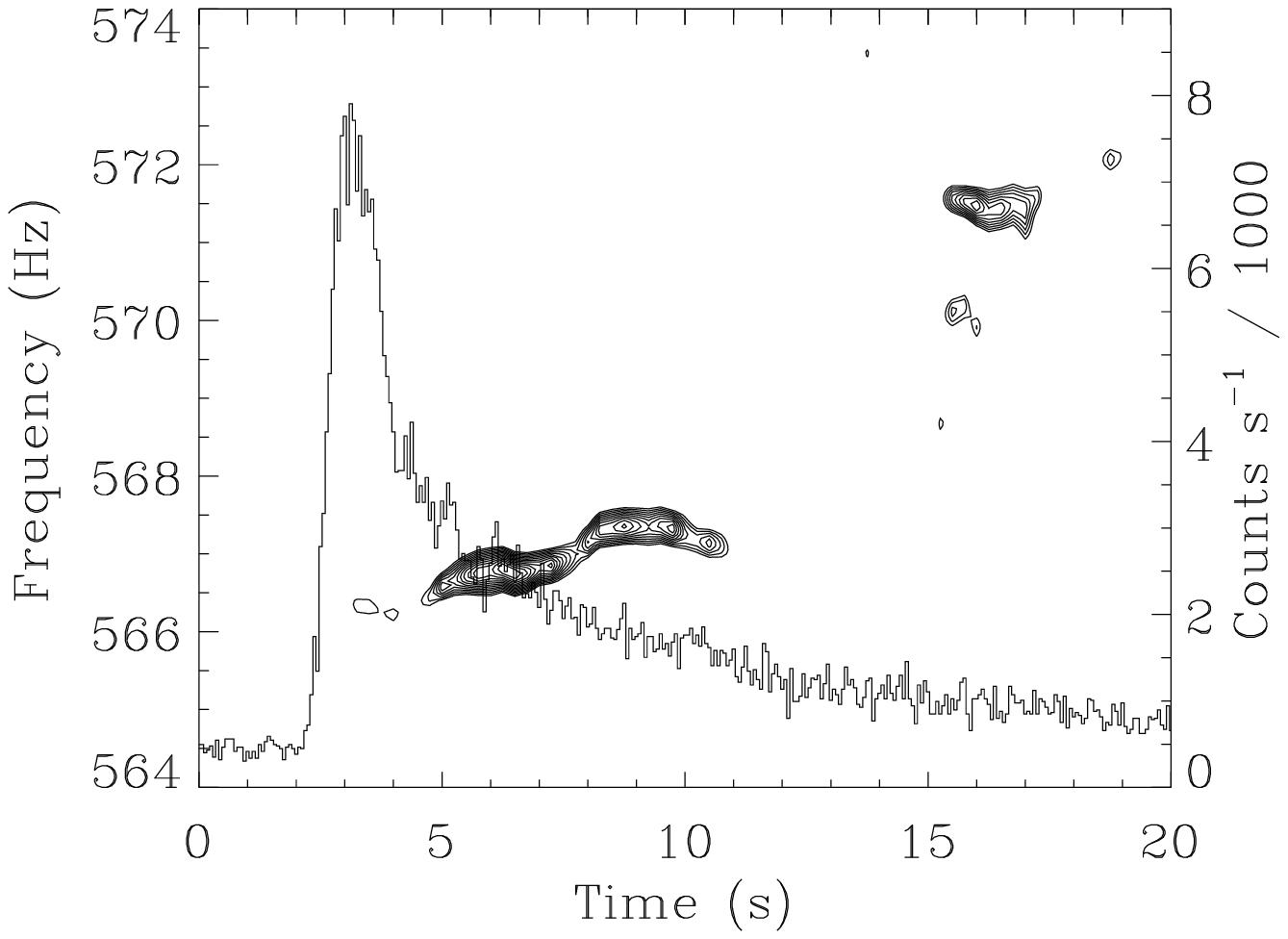,width=8.5cm}\\
\psfig{figure=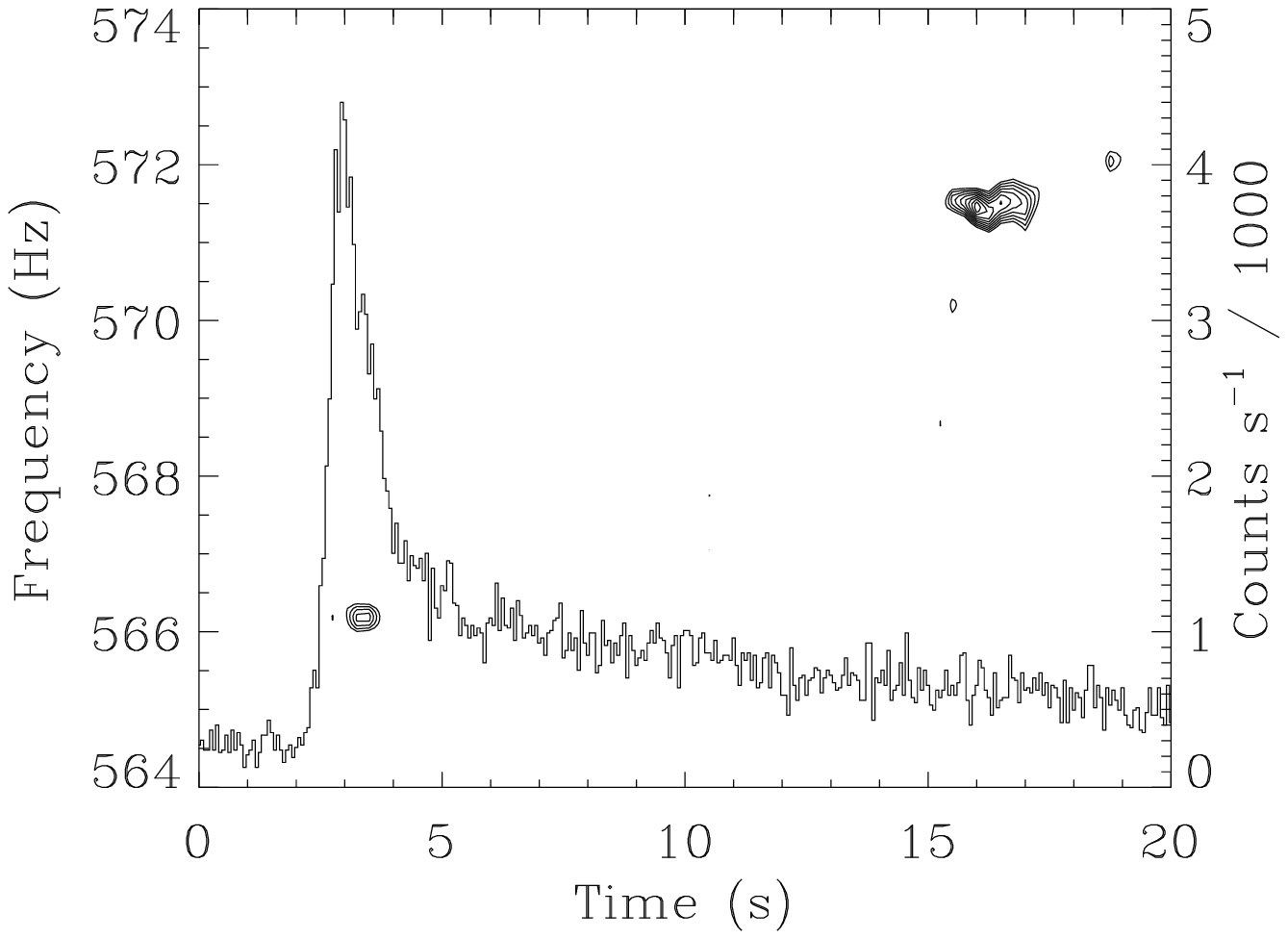,width=8.5cm}\\
\psfig{figure=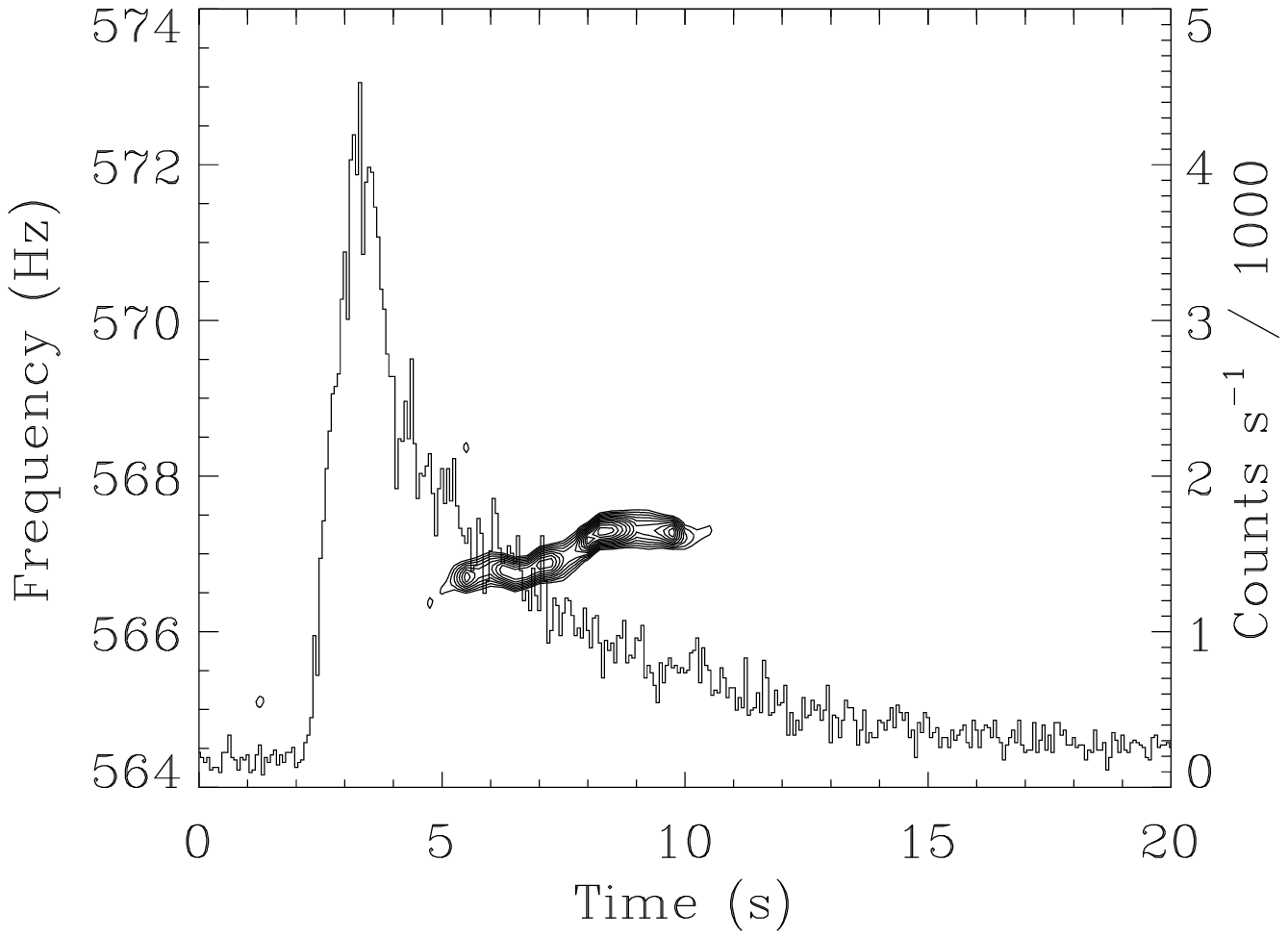,width=8.5cm}
\end{tabular}
\figcaption{The dynamical power spectrum of the burst at 1999 April
14, 11:48 UTC for the energy range 2--22 keV (burst 4; {\it top
panel}), the soft band (2--6 keV; {\it middle panel}), and the hard
band (6--22 keV; {\it bottom panel}). The lowest contour plotted is
$Z^{2}_{1}$ = 14, and the contours go up in steps of 2. The burst
count rate profiles (for 5 PCUs) are over-plotted for the
corresponding energy ranges.  The times given correspond to the start
of the plots.
\label{fig:oscillations_2} }
\end{center}
\end{figure}

The frequency of the oscillation in the tail increases slightly by
about 1 Hz (see Fig.~\ref{fig:oscillations_1}{\it b}), similar to what
has been observed before (e.g., Strohmayer et al. 1996, 1998a;
Markwardt et al. 1999; Muno et al. 2000) in all of the other burst
oscillation sources (although in some bursts a frequency decrease has
been observed; Strohmayer 1999; Miller 1999; Muno et
al. 2000). However, for one burst we observed a frequency increase of
$\sim$5 Hz (Fig.~\ref{fig:oscillations_2} {\it top panel}), the
largest increase so far observed for any system. During this burst the
oscillations have a different energy dependence at different
frequencies: the oscillations are considerably harder when the
frequency is observed near 567 Hz than when it reached 571.5
Hz. Usually the burst oscillations are hard (e.g., Smith et al. 1997;
Strohmayer et al. 1997b; Muno et al. 2000) and the 571.5 Hz signal is
the first clear detection for which the burst oscillations are
soft. The physical mechanism behind this sudden change in hardness is
unclear, although it might be related to the softening of the burst
spectrum in the decay phase of the burst.

In the interpretation that the burst oscillations are due to the
rotation of the neutron star, the evolution of the frequency of the
oscillations (usually an increase of the frequency) observed in the
tail of the bursts has been explained by assuming that the bursting
layer on the neutron star surface expands by several tens of meters
(e.g., Strohmayer et al.  1997; Cumming \& Bildsten 2000). This height
increase causes the layer to slow down and the frequency of the
oscillations to decrease, only to spin up again when the layer settles
back on the neutron star surface. To explain the large (5 Hz) increase
in frequency we observed for X 1658--298, a simple estimate indicates
that the burning layer would have to expand by about 40--45 m. Cumming
\& Bildsten (2000) recently investigated the amount of hydrostatic
expansion expected from bursts over a range of conditions. They find
that $\Delta z(90\%)$, the increase in the height of the layer which
contains 90 \% of the mass, can be as large as about 40 m, so it may
be possible to account for such a change in the frequency.  However,
it is unclear if this 'expanding bursting layer' model can account for
the erratic increase of the frequency during this particular burst.
It is possible that the asymptotic frequency, assumed to be very close
to the spin frequency, is actually a few Hertz lower than the true
spin frequency. This might be the case not only for X 1658--298 but
for all the other sources as well. It is however not clear why the
large frequency increase is only observed for one burst for X
1658--298 and not for the others bursts in this source or for the
bursts in the other sources. In order to study this behavior in more
detail additional bursts are needed which show a similar large
frequency increase.

Besides the five bursts which exhibit the very significant NCOs, we
found a significant detection (although at lower significance) of the
NCOs in one other burst (burst 9) but only non-significant indications
in the other bursts with rms amplitude upper limits significantly
lower than the NCOs strengths (see Tab.~\ref{tab:oscillations}).  Four
of the bursts without NCOs occurred during orbital phase interval
0.75--0.02 (as determined using the ephemeris provided by Wachter et
al. 2000; including the burst during an eclipse).  The absence of NCOs
might be due to the fact that during this orbital phase interval the
source exhibits strong dipping behavior which is most likely caused by
matter coming in to the line of sight. This matter (partly) obscures
our view of the inner system and will also attenuate the NCOs.  The
density of the obscuring matter is highly variable and if the density
decreases significantly, it is possible that the NCOs can become
undetectable, similar to what we observe.

However, the above simple picture does not hold for all bursts.  Three
bursts outside the dipping phase (burst 6, 7, and 13; orbital phase of
0.15, 0.61, and 0.42) do not exhibit these oscillations.  The reason
why these bursts do not show NCOs might be because they behaved
differently from the bursts with oscillations in other respects.
Burst 6 was only about half as bright as the other bursts indicating
that the physical process behind the lower flux of this burst might
also have inhibited the oscillations to be produced.  The burst
profiles of burst 7 and 13 were different from that of the other
bursts. Burst 7 exhibited an excess of burst flux several (2 to 4)
seconds after the peak of the burst, just when oscillations are
expected to be present (as judged from the other bursts which do have
NCOs) and the rise phase of burst 13 last 2 to 4 times as long ($\sim$
4 seconds) as the rise phases of the bursts showing NCOs (typically
1--2 seconds).  The physical processes involved in producing these
unusual burst profiles might also have caused the inhibition of the
NCO mechanism.

Although the above picture can explain why NCOs might not have been
present during all bursts which occur outside the dipping episodes, it
cannot explain the two bursts at phase 0.86 (burst 9) and 0.95 (burst
3) which do show NCOs.  The NCOs in burst 3 can be explained by
assuming that although the source has not entered the eclipsing phase
yet, the dipping activity has already subsided and we are able to
observe the inner system again. This is consistent with the absence of
any clear dipping activity in the persistent emission before and after
this burst (but before the eclipse). However, the NCOs in burst 9
cannot be explained in this way. From the light curve it is clear that
this burst happened during a period of extremely heavy dipping
activity, which should have attenuated the oscillations. Instead the
NCOs we observe in the tail of this burst are even stronger than the
oscillations in the tail observed for other bursts.  If we assume that
the oscillations are indeed partly attenuated then these oscillations
should have been even stronger. It is unclear what causes these strong
oscillations in burst 9 and why they are not strongly attenuated.

X 1658--298 is the first X-ray dipping and eclipsing binary for which
NCOs have been discovered and in this system the binary inclination is
well constrained ($i \sim$ 75\degr--80\degr) compared to the other
systems. Because the other systems do not exhibit X-ray dips their
inclination is lower than $\sim$70\degr~and most likely the range of
inclination angles for these systems is large (based on statistical
grounds). It seems likely that system inclination is an important
factor in determining whether a source will show burst
oscillations. In order for rotational modulation to be effective the
line of sight must not be too close to the spin axis of the neutron
star. Since this axis is almost certainly perpendicular to the orbital
plane systems with high inclination have the most favorable geometry
for producing rotational modulation. The fact that burst oscillations
are seen in X 1658--298 is at least consistent with this simple
argument.  Unfortunately, the inclinations of other burst oscillation
sources are not so tightly constrained, so it is not possible at
present to state that inclination is the primary factor in burst
oscillation observability. Indeed, it may be that other factors are as
important as inclination in producing observable burst oscillations.
For example, the magnetic field strength may be important in pooling
fuel at the magnetic poles, as suggested by Miller (1999) in the
context of 4U 1636-53, or the oscillations might only be detectable
above certain values of the mass accretion rate as shown for KS
1731--260 (Muno et al. 2000) and 4U 1728--34 (Franco 2000; van
Straaten et al. 2000). At the moment, however, it is still unclear why
only a few sources exhibit these burst oscillations.

Due to its high inclination and hence the strong constraints on the
binary parameters, X 1658--298 could be an excellent source in order
to try to determine the X-ray mass function from the burst
oscillations as proposed by Strohmayer et al. (1998b).  The bursts
which exhibit the oscillations during the tail of the bursts (four
bursts in our sample of bursts) are the most promising bursts in order
to perform such a measurement if one assumed that the asymptotic
frequency observed is the neutron star spin frequency.  However, the
observation of a burst frequency several Hz higher than the asymptotic
frequency in the other three bursts casts doubt on the validity of
this assumption.

\acknowledgments

This work was supported by NASA through Chandra Postdoctoral
Fellowship grant number PF9-10010 awarded by CXC, which is operated by
SAO for NASA under contract NAS8-39073.  This research has made use of
data obtained through the HEASARC Online Service, provided by the
NASA/GSFC.

\clearpage

\begin{deluxetable}{rllllclclc}
\tabletypesize{\footnotesize}
\rotate
\tablecolumns{10}
\tablewidth{0pt}
\tablecaption{The properties of the X-ray bursts\label{tab:oscillations}}\vspace{-0.4cm}
\tablehead{\vspace{-0.2cm}
\#     & ObsID          & Time of burst     &  Orbital   & PCUs on   & Count rate$^b$      &  Osc.         & Amplitude$^c$    & Single trial$^d$     & Amplitude$^e$\\
       &                & (UTC, 1999)       &  phase$^a$ &           & (counts/s/pcu)      &               & (567 Hz; \% rms) & probability          & (571.5 Hz; \% rms) }
\startdata
 1     & 40036-10-01-00 & Apr 6 12:12:46    &  0.80      & 0,1,2,3,4 & 2120                &  No           &  $<$4.6          & --                   & $<$ 6.9 \\ [-.2cm]
 2     & 40050-04-01-00 & Apr 9 14:46:30    &  0.28      & 0,1,2,3,4 & 2060                &  Rise         &  19.6            & $1.1\times10^{-10}$  & $<$ 9.1 \\[-.2cm]
 3     & 40050-04-02-00 & Apr 10 9:48:33    &  0.95      & 0,1,2,3,4 & 2190                &  Tail         &   8.5            & $1.5\times10^{-9}$   & $<$ 4.7 \\[-.2cm]
 4     & 40050-04-04-00 & Apr 14 11:47:52   &  0.73      & 0,1,2,3,4 & 1580                &  Tail         &   8.7            & $2.2\times10^{-9}$   & 12.3\pp1.5    \\[-.2cm]
 5     & 40050-04-08-00 & Apr 18 16:37:09   &  0.90      & 0,2,3,4   & 2400                &  No           &  $<$7.4          & --                   & $<$ 9.2 \\[-.2cm]
 6     & 40050-04-08-00 & Apr 18 18:26:14   &  0.15      & 0,2,3,4   &  950                &  No           &  $<$6.1          & --                   & $<$ 14.5 \\[-.2cm]
 7     & 40050-04-10-00 & Apr 20 16:24:19   &  0.61      & 0,1,2,3,4 & 1840                &  No           &  $<$4.9          & --                   & $<$ 11.1 \\[-.2cm]
 8     & 40050-04-11-00 & Apr 21 11:44:53   &  0.33      & 0,2,3,4   & 1800                &  Tail         &   8.0            & $1.3\times10^{-7}$   & $<$  7.7 \\[-.2cm]
 9     & 40050-04-13-00 & Apr 24 14:42:08   &  0.86      & 0,1,2,3   &  275                &  Tail         &  12.6            & $1.0\times10^{-6}$   & $<$  9.1 \\[-.2cm]
10     & 40050-04-16-00 & Apr 29 14:50:32   &  0.75      & 0,2,4     & 1380                &  No           &   $<$7.0         & --                   & $<$ 10.9 \\[-.2cm]
11     & 40050-04-20-00 & May 5 11:08:47    &  0.46      & 0,1,2,3   & 1385                &  Rise         &  12.0            & $1.8\times10^{-7}$   & $<$  8.9 \\[-.2cm]
12     & 40050-04-21-00 & May 9 17:23:12    &  0.83      & 0,1,2,3   & 1625                &  No           &  $<$9.2          & --                   & $<$ 20.9 \\[-.2cm]
13     & 40050-04-23-00 & Jun 2 7:41:29     &  0.42      & 0,2,3,4   & 1520                &  No           &  $<$4.9          & --                   & $<$  8.9 \\[-.2cm]
14     & 50410-01-05-00 & Aug 8 9:32:22$^f$ &  0.02      & 0,2,3,4   &  575                &  No           &  $<$14.1         & --                   & $<$  8.6 \\
\enddata

\tablenotetext{a}{As determined using the ephemeris of Wachter et
al. 2000. Mid-eclipse time is phase 0.0}

\tablenotetext{b}{The peak count rate for the 2--60 keV {\it RXTE}/PCA energy
range}\vspace{-0.4cm}

\tablenotetext{c}{For 2--60 keV. The amplitude is the maximum
amplitude of the oscillations throughout the bursts. The errors on the
amplitudes are between 1\% and 1.5\% rms.  The 90 \% confidence upper
limits are calculated for the beginning of the bursts to 20 s after
the start in the frequency range 564--568 Hz}

\tablenotetext{d}{The single trial probability for chance detections.}

\tablenotetext{e}{For 2--60 keV. The amplitude for burst 4 is the
maximum amplitude of the oscillations.  The 90 \% confidence upper
limits are calculated for the peak of the bursts to 20 s after the
start in the frequency range 570--573 Hz. }

\tablenotetext{f}{In the year 2000. Approximate start time of this
burst; it started when the source was still eclipsed by the companion
star.}

\end{deluxetable}

\end{document}